\documentclass[prb,twocolumn,showpacs,floatfix]{revtex4-1}
\usepackage{graphicx}
\usepackage{color}
\usepackage{amssymb}
\usepackage{amsmath}
\usepackage{xspace}
\usepackage{url}
\usepackage{subfigure}
\usepackage{float}

\let \Re \relax
\DeclareMathOperator{\Re}{Re}
\let \Im \relax
\DeclareMathOperator{\Im}{Im}
\newcommand{\q}{{\bf q}}
\newcommand{\s}{{\phantom{\dagger}}}

\newcommand{\beps}{\bar\varepsilon}
\newcommand{\teps}{\tilde\varepsilon}
\newcommand{\inda}{{{\bf k}\sigma}}
\newcommand{\indo}{{{\bf k}'\,-\sigma}}
\newcommand{\wxo}{\omega_X^{-\sigma,\sigma}}
\newcommand{\llangle}{\langle\langle}
\newcommand{\rrangle}{\rangle\rangle}

\begin{document} 
\title{Electron-hole pair condensation at the semimetal-semiconductor transition:
\\
a BCS-BEC crossover scenario}

%
\author{B. Zenker$^1$, D. Ihle$^2$, F. X. Bronold$^1$, and H. Fehske$^{1}$}
\affiliation{$^1$Institut f{\"u}r Physik,
            Ernst-Moritz-Arndt-Universit{\"a}t Greifswald,
             D-17487 Greifswald, Germany \\
             $^2$Institut f{\"u}r Theoretische Physik, Universit{\"a}t Leipzig, D-04109 Leipzig, Germany}

\date{\today}
\begin{abstract}
We act on the suggestion that an excitonic insulator state might 
separate---at very low temperatures---a semimetal from a semiconductor
and ask for the nature of these transitions. Based on the analysis of
electron-hole pairing in the extended Falicov-Kimball model,
we show that tuning the Coulomb attraction between both species,
a continuous crossover between a BCS-like transition of Cooper-type pairs
and a Bose-Einstein condensation of preformed tightly-bound excitons
might be achieved in a solid-state system. The precursor of this
crossover in the normal state might cause the transport anomalies 
observed in several strongly correlated 
mixed-valence compounds.  
\end{abstract}
\maketitle
The challenging suggestion of electron-hole pair condensation in thermal equilibrium
into the excitonic insulator (EI) phase at the semimetal (SM) to semiconductor (SC) transition~\cite{HR68b},
where the SM--EI transition may be described in analogy with BCS theory of superconductivity
and the SC--EI transition is discussed in terms of a Bose-Einstein condensation (BEC) of preformed
excitons~\cite{BF06,IPBBF08,SEO11}, is of topical interest. This is due to the growing 
amount of experimental data on materials which are candidates for the realization of the EI, where different
situations with respect to the SM/SC--EI transition are given. For
example, in the rare-earth chalcogenide $\rm TmSe_{0.45}Te_{0.55}$, 
that is, an intermediate-valent SC, the pressure-induced resistivity anomaly at
low temperatures was ascribed to exciton formation and a subsequent SC--EI transition~\cite{NW90,BSW91,Wa01,WBM04}. 
An EI state in semiconducting  $\rm Ta_2NiSe_5$  
was recently probed by photoemission~\cite{Wa09}.
On the other hand, in the layered transition-metal 
dichalcogenide 1$T$--$\rm TiSe_2$,
which is a SM, a BCS-like electron-hole pairing was considered as 
the driving force for the periodic lattice distortion~\cite{MSGMDCMBTBA10}.
Here evidence suggests electron-hole `Cooper-pair' fluctuations above the
SM-EI transition temperature. A BCS-like electron-hole
pair condensation was also studied for graphene bilayers~\cite{LS08b}. 
In this system a BCS-BEC crossover might be realized  
by a magnetic field that creates a gap 
and magneto-excitons which may condense. From a
theoretical point of view, one of the main issues in this field is the
better understanding and a detailed description of the normal phase above the
SM/SC--EI transition, especially of the electron--hole pair
fluctuations and of the BCS--BEC crossover scenario~\cite{Le80}
that characterizes the EI instability and has not been observed in a solid
so far.

In this Rapid Communication we address this topic
and the mechanisms behind 
in terms of a minimal two-band model, the so-called 
extended Falicov-Kimball model (EFKM)~\cite{Ba02b,Fa08,IPBBF08}
which covers direct $c$- and $f$-band hopping and admits 
the pairing of $c$ electrons with $f$ holes via a 
strongly screened Coulomb interaction. 
Thereby we focus on the normal phase that surrounds the EI
and look for precursor effects in the electron-hole pair
susceptibility. In particular we analyze the nature
of the electron-hole bound states and determine their number 
and spectral weight. 
We are able to show how the 
normal-state to EI transition changes from 
BCS to BEC when the SM gives way to the SC.    
    
Representing the orbital flavor of the $f$, $c$ electrons by the pseudospin
$\sigma=\uparrow,\downarrow$, the EFKM takes the form 
\begin{equation}
H=\sum_{{\bf k},\sigma} \varepsilon_{{\bf k}\sigma} n_{{\bf k}\sigma} + U\sum_i n_{i\uparrow} n_{i\downarrow}\,.
\label{EFKM}
\end{equation}
Equation~\eqref{EFKM} constitutes a generalized Hubbard model with on-site Coulomb 
interaction $U$ and spin-dependent band energies
$\varepsilon_{{\bf k}\sigma}=E_\sigma - t_\sigma \gamma_{\bf k}-\mu$, 
where $E_\sigma$ defines the band-center of the $\sigma$-band, 
$t_\sigma$ denotes the nearest-neighbor hopping amplitude 
on a $D$-dimensional hypercubic lattice, 
$\gamma_{\bf k}=2\sum_{d=1}^D \cos k_d$, and $\mu$ is the 
chemical potential. For $E_\uparrow < E_\downarrow$ and  
$t_\uparrow t_\downarrow<0$ ($t_\uparrow t_\downarrow>0$) a direct 
(indirect) band gap might appear. The $\sigma$-electron density 
is given by \mbox{$n_\sigma = \frac{1}{N}\sum_{\bf k} \langle n_{{\bf k}\sigma} \rangle =\frac{1}{N} \sum_{\bf k} \langle c_{{\bf k}\sigma}^\dagger c_{{\bf k}\sigma}^\s \rangle$}, and we require $n_\uparrow+n_\downarrow=1$ for the 
half-filled band case.

The EI low-temperature phase of the EFKM is characterized by a non-vanishing 
order parameter $\Delta=\frac{U}{N}\sum_{\bf k} \langle c_{{\bf k}\downarrow}^\dagger c_{{\bf k}\uparrow}^\s\rangle$
(in case of a direct band gap)~\cite{IPBBF08,Fa08,ZIBF10}. 
Describing a condensate
of electron-hole pairs (excitons), $\Delta$  obeys a gap equation with anomalous
Green's functions 
involved~\cite{BF06,MSGMDCMBTBA10,SEO11}. 
From this the transition temperature $T_{\rm EI}(U)$
can be determined. In what follows we scrutinize the existence of 
excitonic bound states above $T_{\rm EI}$ where $\Delta =0$.
To this end we analyze the susceptibility 
$\chi_\q^{\uparrow,\downarrow}(\omega)=\llangle b_\q^\s; b_\q^\dagger \rrangle_\omega$,
with $b_\q^\dagger =\frac{1}{\sqrt{N}}\sum_{\bf k} c_{{\bf k}+\q\downarrow}^\dagger c_{{\bf k}\uparrow}^\s$ creating an electron-hole excitation with momentum
${\bf q}$ in the SM and SC high-temperature phases. 
The pole of $\chi_\q^{-\sigma,\sigma}(\omega)$, $\omega_X(\q)=\omega_X^{\uparrow,\downarrow}(\q)=-\omega_X^{\downarrow,\uparrow}(\q)$, calculated in ladder approximation, describes an exciton, provided that $0<\omega_X(\q)<\omega_C(\q)$. Here $\omega_C(\q)={\rm min}_{\bf k}(\teps_{{\bf k}+\q\downarrow}-\teps_{{\bf k}\uparrow})$ 
is the lower bound of the electron-hole excitations and $\teps_\inda$
denotes the renormalized band structure. The binding energy of the exciton 
is $E_B^X(\q)=\omega_C(\q)-\omega_X(\q)$. Outside the 
electron-hole continuum the imaginary part of 
$\chi_\q^{-\sigma,\sigma}$ is 
\begin{equation}
\Im\chi_{\bf q}^{-\sigma,\sigma}(\omega) =-\pi\, Z(\omega_X,{\bf q})\, \delta(\omega-\wxo) ,
\label{ImChi1}
\end{equation}
where
\begin{equation}
Z(\omega_X,{\bf q})=\left[ \frac{U^2}{N}\sum_{\bf k}\frac{f(\teps_{{\bf k}\uparrow})-f(\teps_{{\bf k}+{\bf q}\downarrow})}{(\omega_X+\teps_{{\bf k}\uparrow}-\teps_{{\bf k}+{\bf q}\downarrow})^2} \right]^{-1} 
\label{Z}
\end{equation}
gives the spectral weight of the excitonic quasiparticle.

To determine the chemical potential 
including self-energy effects, we expand
the imaginary part of $G_\inda(\omega)=[\omega-\bar{\varepsilon}_{{\bf k}\sigma}-\Sigma_\inda(\omega)]^{-1}$, where $\bar{\varepsilon}_{{\bf k}\sigma}=\varepsilon_{{\bf k}\sigma}+U n_{-\sigma}$, for small damping. 
The self-energy is obtained by the Green's function projection technique~\cite{Pl11}: 
\begin{eqnarray}
\Sigma_\inda(\omega) &=& -(U^2/N \pi^2)\sum_{{\bf k}'} \int d\bar\omega \int d\omega'\,\frac{\left[f(\omega')+p(\omega'-\bar\omega)\right]}{\omega-\bar\omega}
\nonumber \\
&&\times \Im \chi_{{\bf k}-{\bf k}'}^{-\sigma,\sigma}(\bar\omega-\omega') \Im G_\indo(\omega')\end{eqnarray}
 with  $f(\omega)=\left[e^{\beta\omega}+1\right]^{-1}$, $p(\omega)=[e^{\beta \omega}-1]^{-1}$. 
Considering the parameter region near the SM/SC-EI transition, where the 
dominant weight of the electron-hole spectral function is suggested to arise 
from the bound state as compared with the electron-hole continuum, in the 
self-energy calculation we take into account only the excitonic quasiparticle 
contribution given by Eq.~\eqref{ImChi1}.  
Then the $\sigma$-electron density can be decomposed into a part of nearly-free electrons (with 
renormalized dispersion) and a term $\propto \Im\Sigma_\inda (\omega)$ that comprises the electron-bound states as well as the reaction of the $\sigma$-electrons to the 
existence of excitons. Denoting the latter contribution as the correlation part, 
we have $n_\sigma=n_\sigma^{\rm nf}+n_\sigma^{\rm corr}$, and find  
$n_\sigma^{\rm nf}=\frac{1}{N}\sum_{\bf k} f(\teps_\inda)$,
where $\teps_\inda = \bar{\varepsilon}_{{\bf k}\sigma}+\Re\Sigma_\inda (\omega)\big|_{\omega=\teps_\inda}$.   
It turns out that $n_\uparrow^{\rm corr}=-n_\downarrow^{\rm corr}$. 
Hence, the chemical potential can be obtained by using solely 
the nearly-free part of the particle densities, i.e., 
$n_\uparrow+n_\downarrow=n_\uparrow^{\rm nf}+n_\downarrow^{\rm nf}=1$ 
(cf. also Ref.~\onlinecite{BF06}).

The number of excitons with center-of-mass momentum $\q$ results in
\begin{equation}
N_X(\q)=\langle b_{\bf q}^\dagger\,b_{\bf q}^\s \rangle \big|_{\omega_X} = Z(\omega_X,\q)\;p(\omega_X)\,,
\label{Xnumbr}
\end{equation}
leading to the total exciton density $n_X=\frac{1}{N} \sum_\q N_X(\q)$.
To characterize the composition of the normal phase, we introduce the 
bound-state fractions
$\Gamma=n_X/(n_X+n_\downarrow^{\rm nf})$ and $\Gamma_0=N_X(0)/N(n_X+n_\downarrow^{\rm nf})$.

To simplify the numerical analysis, we discard the 
band renormalization of the $\sigma$-electrons by the 
excitons, i.e., we neglect the term $\propto {\rm Re} \Sigma$ 
in $\teps_\inda$. Then $n_\sigma^{\rm nf}$ contains the Hartree shift 
$Un_{-\sigma}$ only and, inserting the nearly-free part of $G_{{\bf k}\sigma}$ into 
$\Sigma_{{\bf k}\sigma}$
$\chi_\q^{-\sigma,\sigma}$ becomes the RPA result. 
Since the ground-state phase diagram of the EFKM 
is similar in 2D and 3D~\cite{Ba02b,Fa08}, 
and we are primary interested in the normal-state properties 
for $T>T_{\rm EI}$, we consider the 2D case hereafter.
To model an intermediate-valence situation we choose 
$E_\uparrow=-2.4$, $E_\downarrow=0$, $t_\uparrow=-0.8$ without loss of 
generality, and take $t_\downarrow=1$ as energy unit. 

\begin{figure}[t]
\centering
\includegraphics[width=0.8\linewidth]{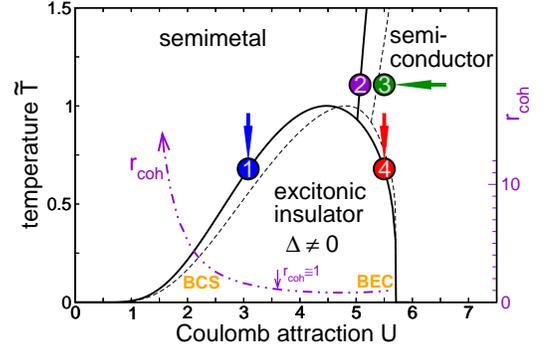}
\caption{(Color online) EI formation at the SM-SC transition in the 
2D EFKM. The phase diagram is calculated for a band splitting 
$E_\uparrow-E_\downarrow=-2.4$, and band asymmetry 
$t_\uparrow/t_\downarrow=-0.8$, where the temperature $T$ is 
scaled to the maximum critical temperature  
$\tilde{T}=T/T_{\rm EI}^{\rm max}$
with $T_{\rm EI}^{\rm max}=0.361$ (RPA, solid line) 
and $T_{\rm EI}^{\rm max}=0.256$ 
(slave boson, thin dashed line). The coherence length of the EI 
condensate at $\tilde{T}=0$, $r_{\rm coh}$ 
(as defined in Ref.~\onlinecite{SEO11}), is indicated by the 
dot-dot-dashed line.}
\label{fig1}
\end{figure}

The RPA EFKM phase diagram shown in Fig.~\ref{fig1} describes the 
general scenario at the SM-SC transition, which persists, apart from a 
reduction of the critical temperature, also when the electronic correlations
are treated by the 
more elaborate slave boson approach~\cite{ZIBF10}. 
The SM (SC) has a
gapless (gapful) band structure with a small band overlap (band gap).
The metal-insulator transition is triggered by enlarging the Hartree shift   
upon increasing $U$. The phase boundary to the EI can be obtained
from a BCS-like gap equation, which holds on both SM and SC sides, 
but gives no detailed insight into the nature of this transition. 

\begin{figure}[t]
\subfigure{
\includegraphics[width=\linewidth]{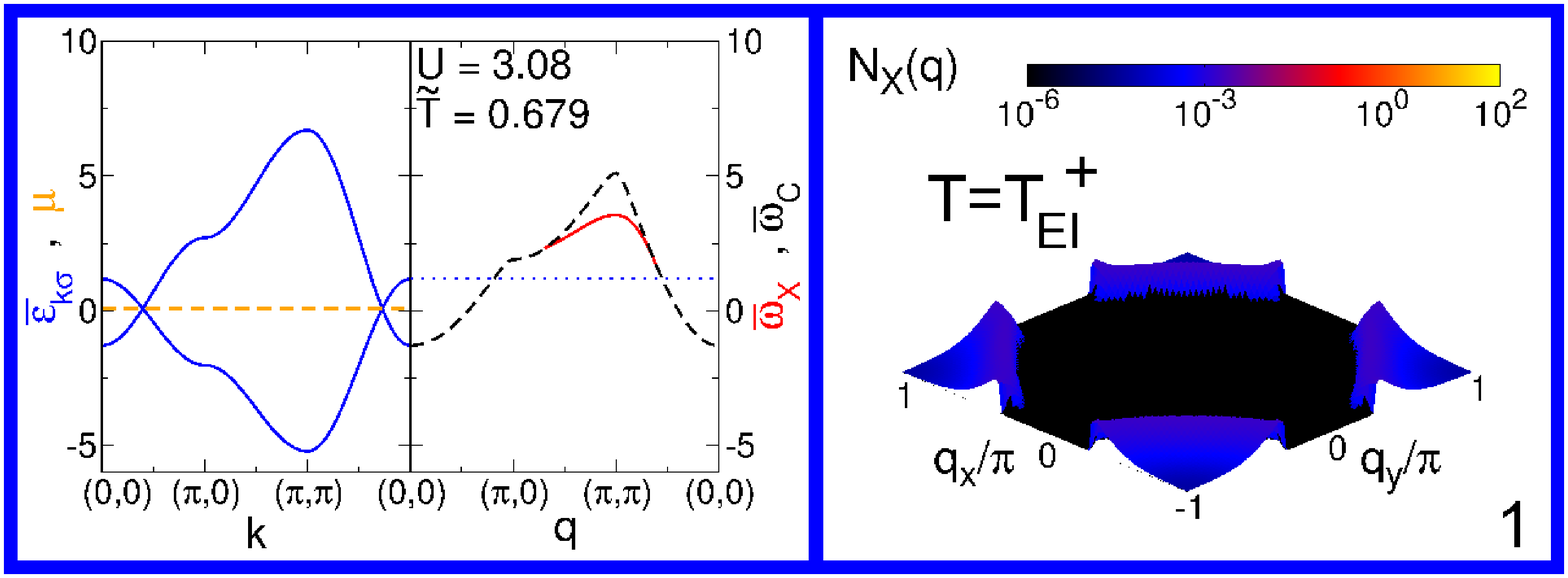}
}
\subfigure{
\includegraphics[width=\linewidth]{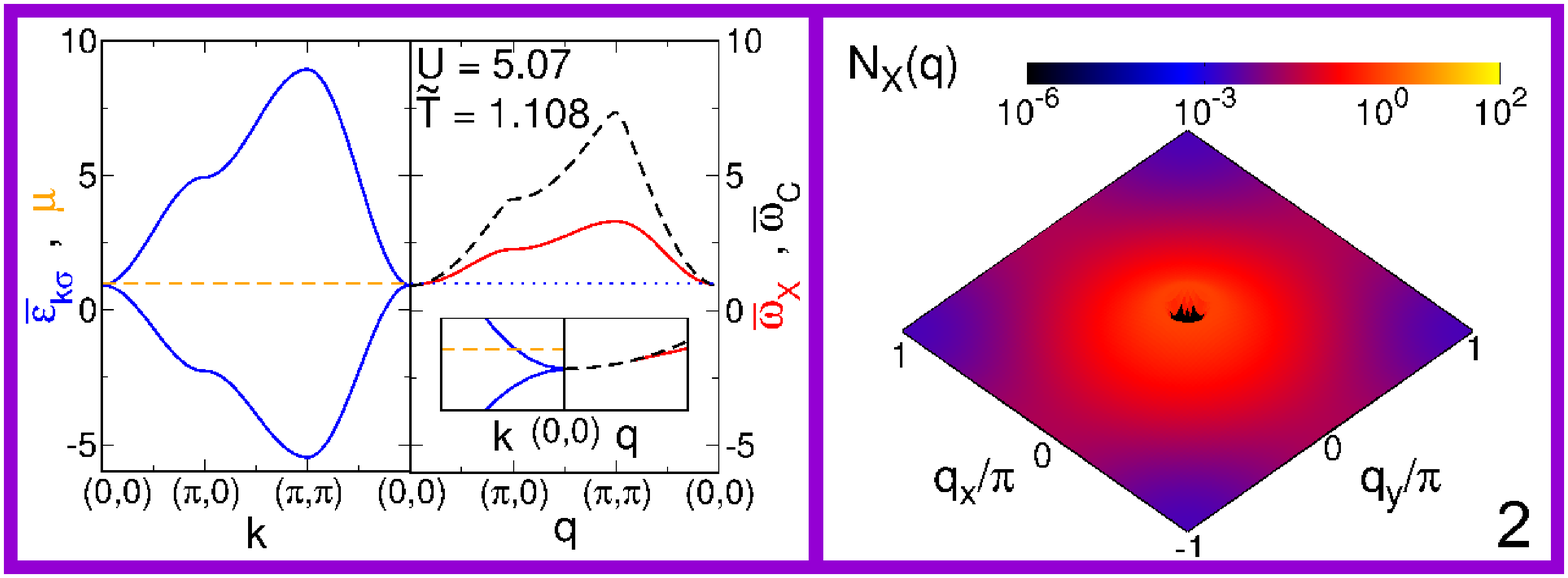}
}
\subfigure{
\includegraphics[width=\linewidth]{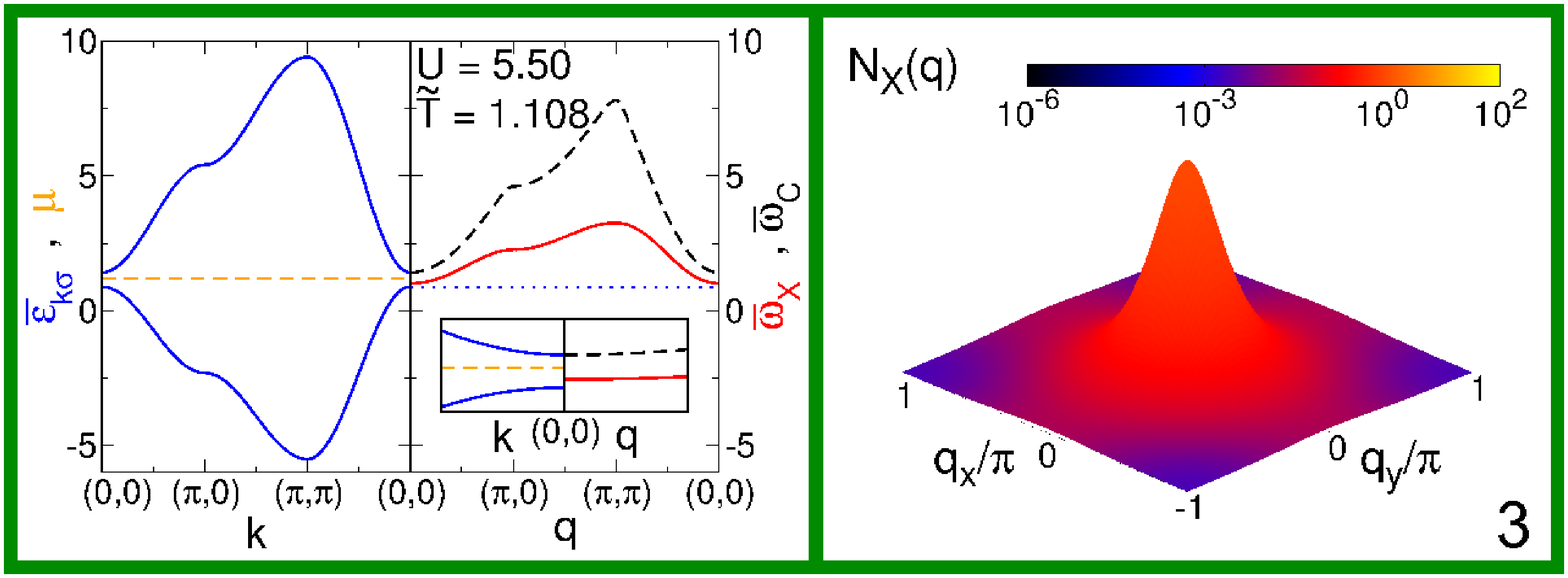}
}
\subfigure{
\includegraphics[width=\linewidth]{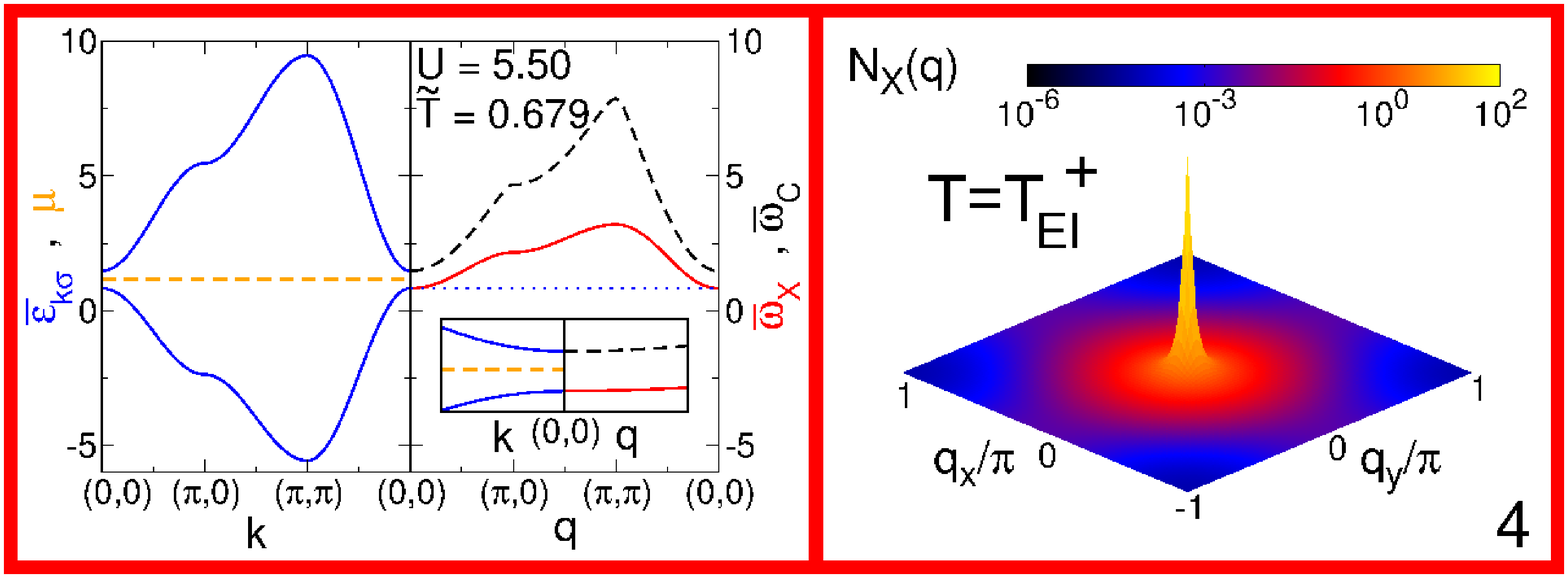}
}
\caption{(Color online) Mean-field band scheme  (left panels) and 
${\bf q}$-resolved exciton numbers (right panels) 
for the points marked in Fig.~\ref{fig1} (1: at the SM-EI transition, 2: at the SM-SC transition, 3: in the SC regime, 4: at the SC-EI transition). The electron dispersion ($\bar{\varepsilon}_{{\bf k}\sigma}$)  is given by the blue solid lines, the chemical potential ($\mu$) by the orange dashed line. Note that the excitonic level ($\omega_X(\q)$, red solid line) and the continuum level ($\omega_C(\q)$, black dashed line) are shifted by the $\uparrow$-band maximum, $\bar\omega_{X/C}= \omega_{X/C}+{\rm max}_{\bf k}(\beps_{{\bf k}\uparrow})$. 
The blue dotted line marks the $\uparrow$-band top.}
\label{fig2}
\end{figure}

\begin{figure}[t]
\subfigure[$U=0.50$,\newline $\widetilde{T}=1.108$]{
\includegraphics[height=0.25\linewidth]{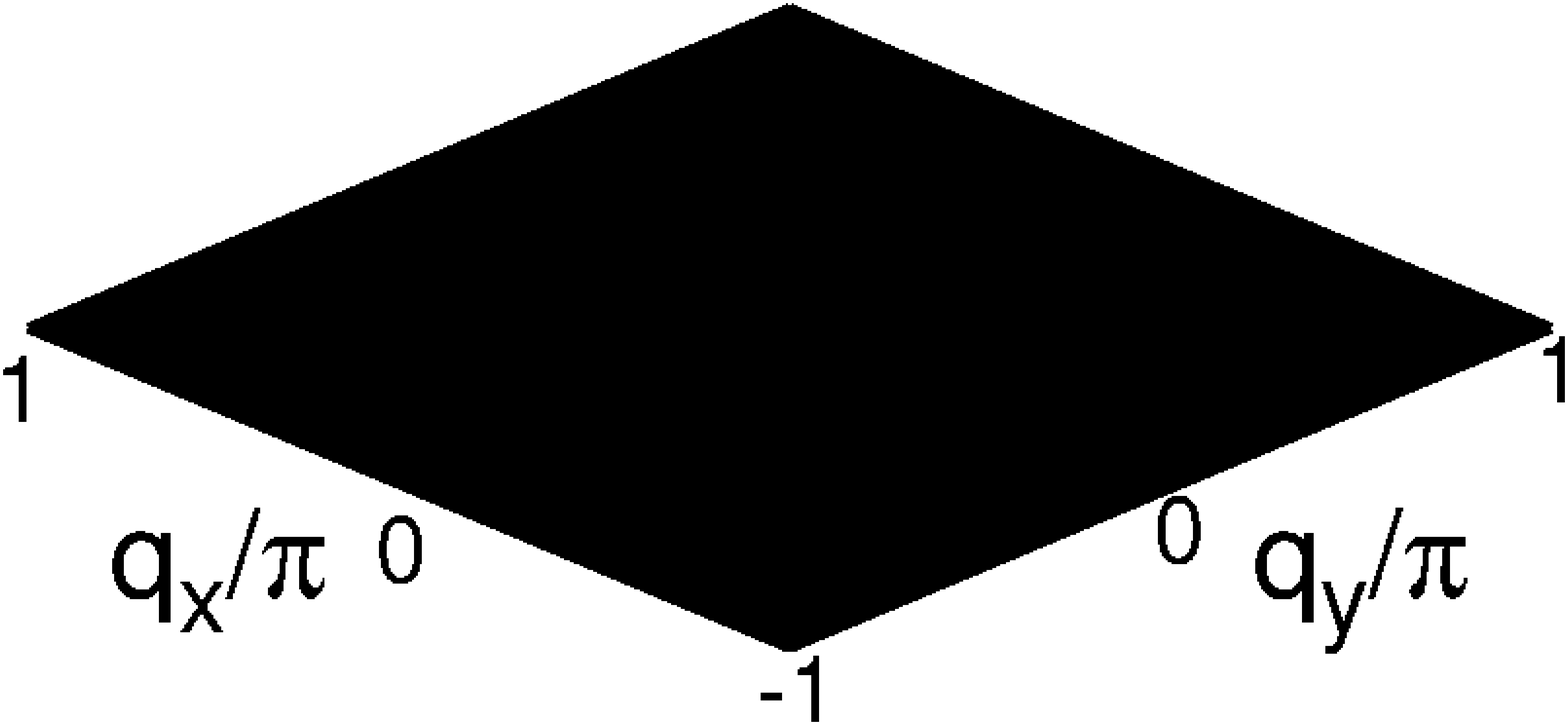}
\label{fig3a}
}
\subfigure[$U=3.08$,\newline $\widetilde{T}=0.679$]{
\includegraphics[height=0.25\linewidth]{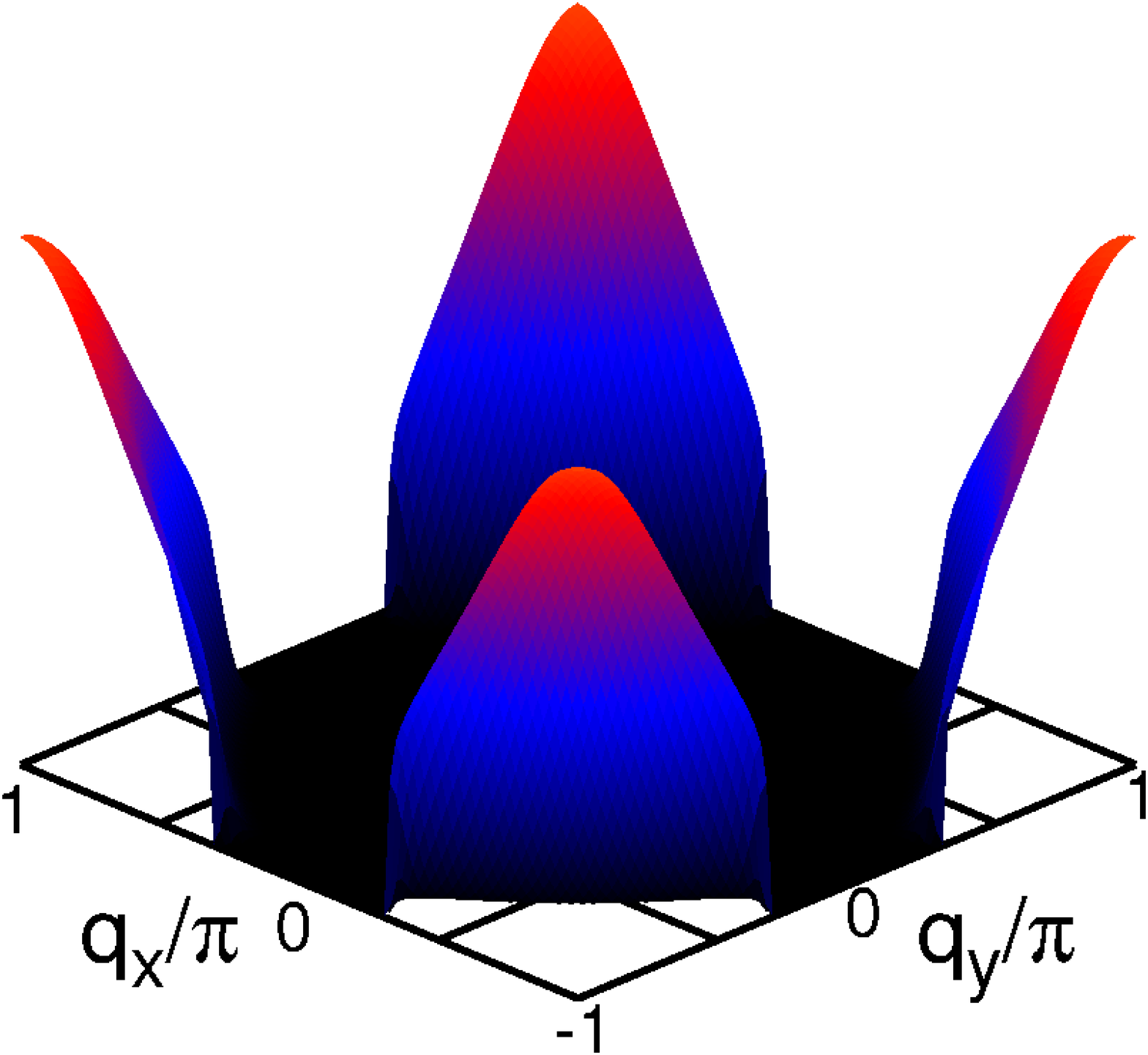}
\label{fig3b}
}
\subfigure[$U=5.07$,\newline $\widetilde{T}=1.108$]{
\includegraphics[height=0.25\linewidth]{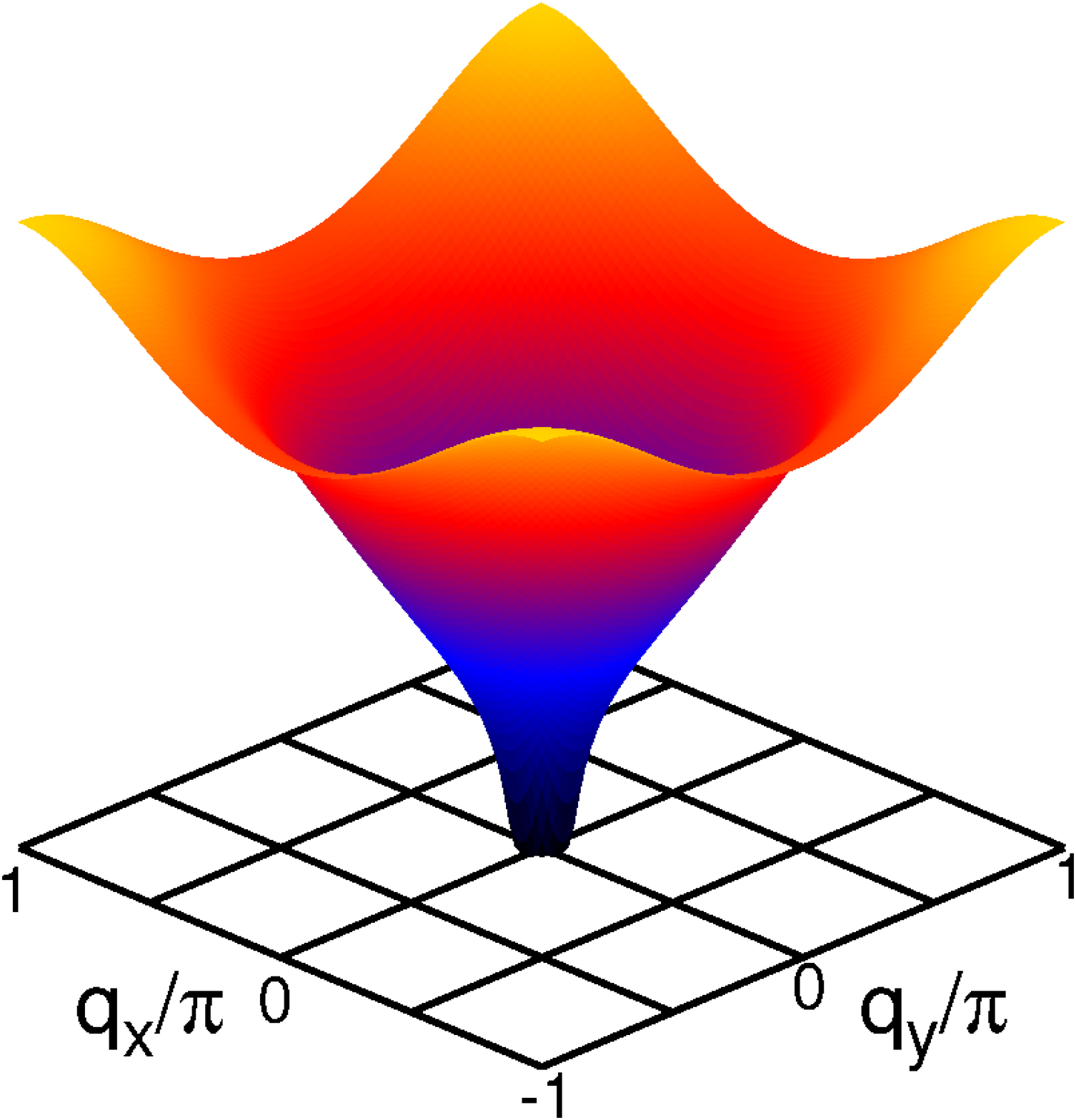}
}
\subfigure{
\includegraphics[height=0.25\linewidth]{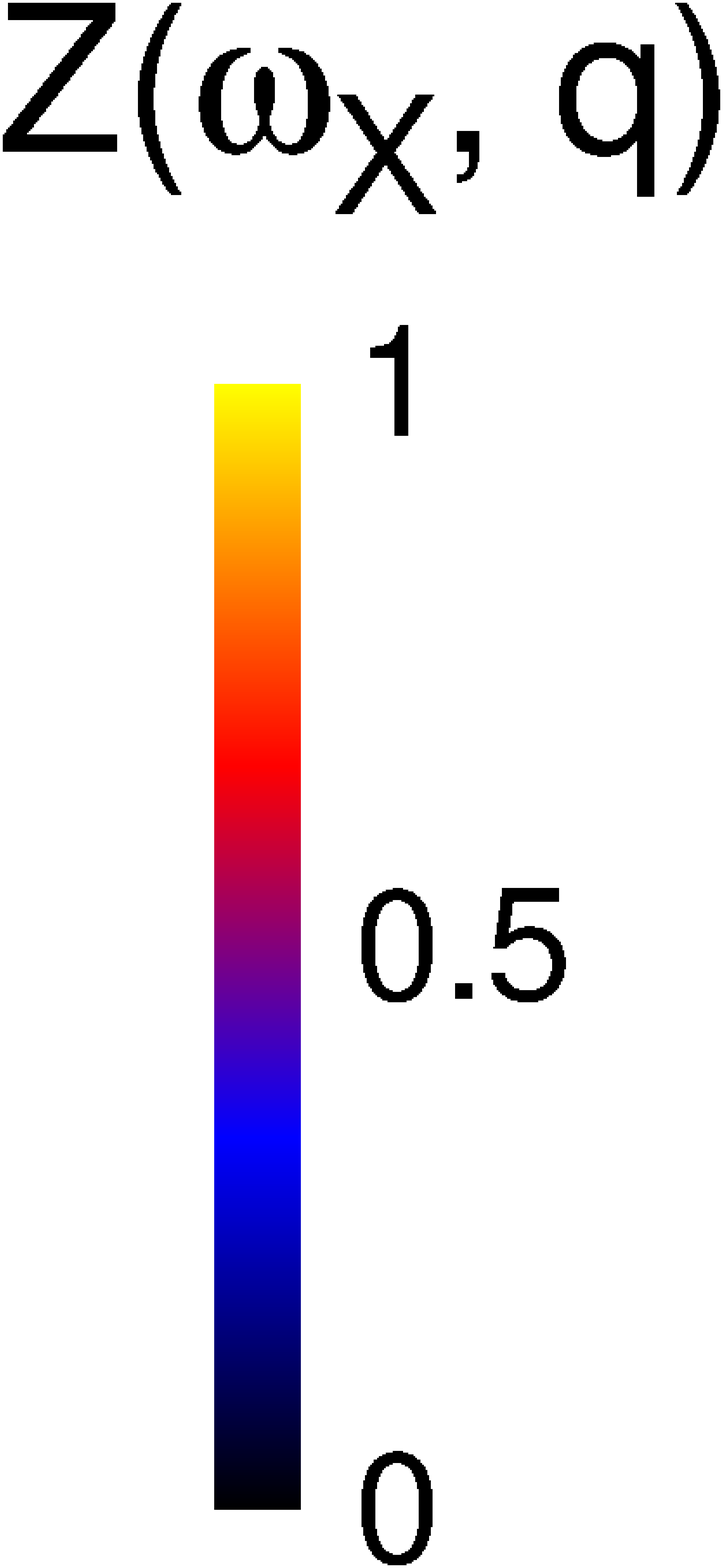}
\label{fig3d}
}
\subfigure[$U=5.50$,\newline $\widetilde{T}=1.108$]{
\includegraphics[height=0.25\linewidth]{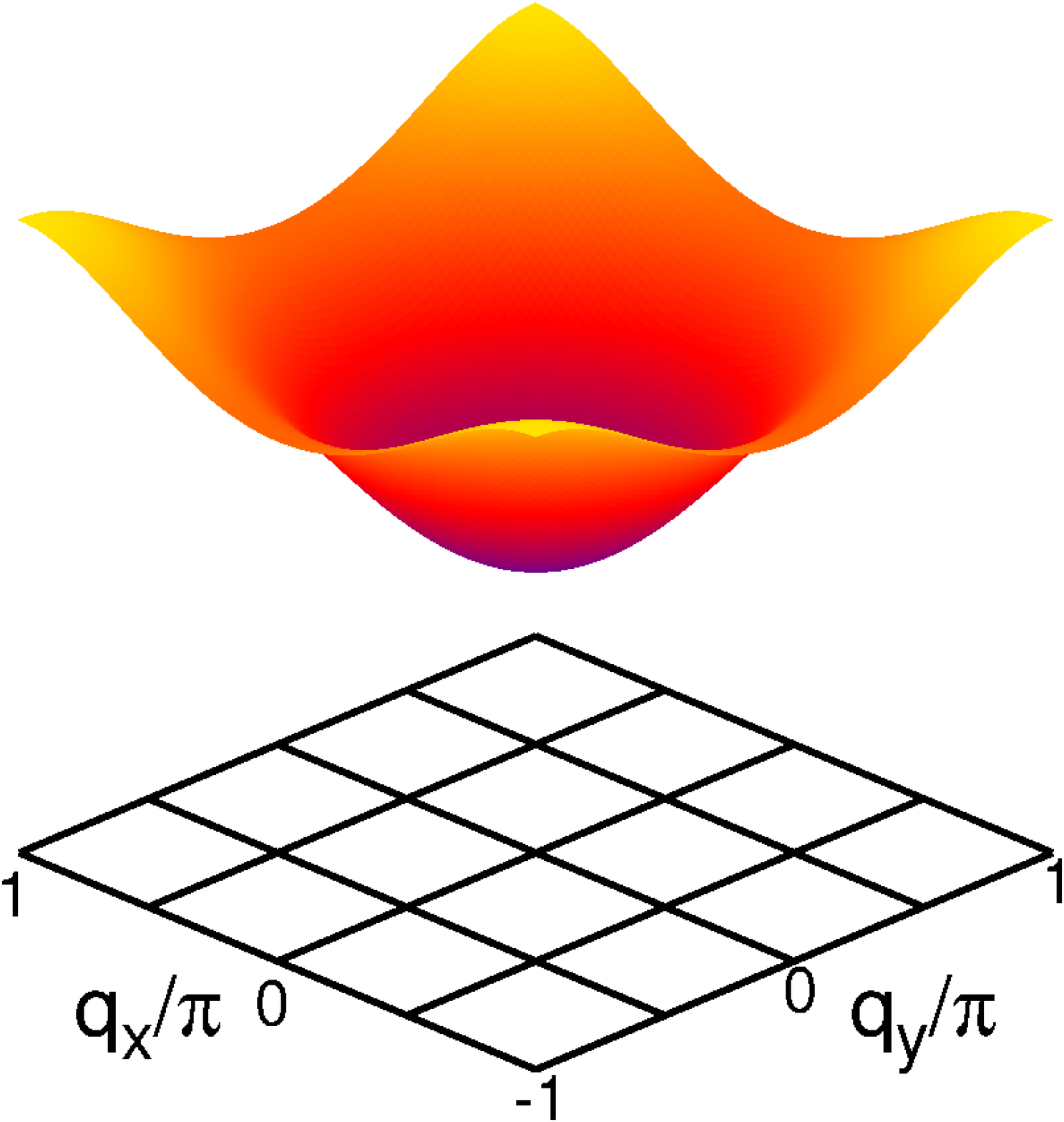}
\label{fig3e}
}
\subfigure[$U=5.50$,\newline $\widetilde{T}=0.679$]{
\includegraphics[height=0.25\linewidth]{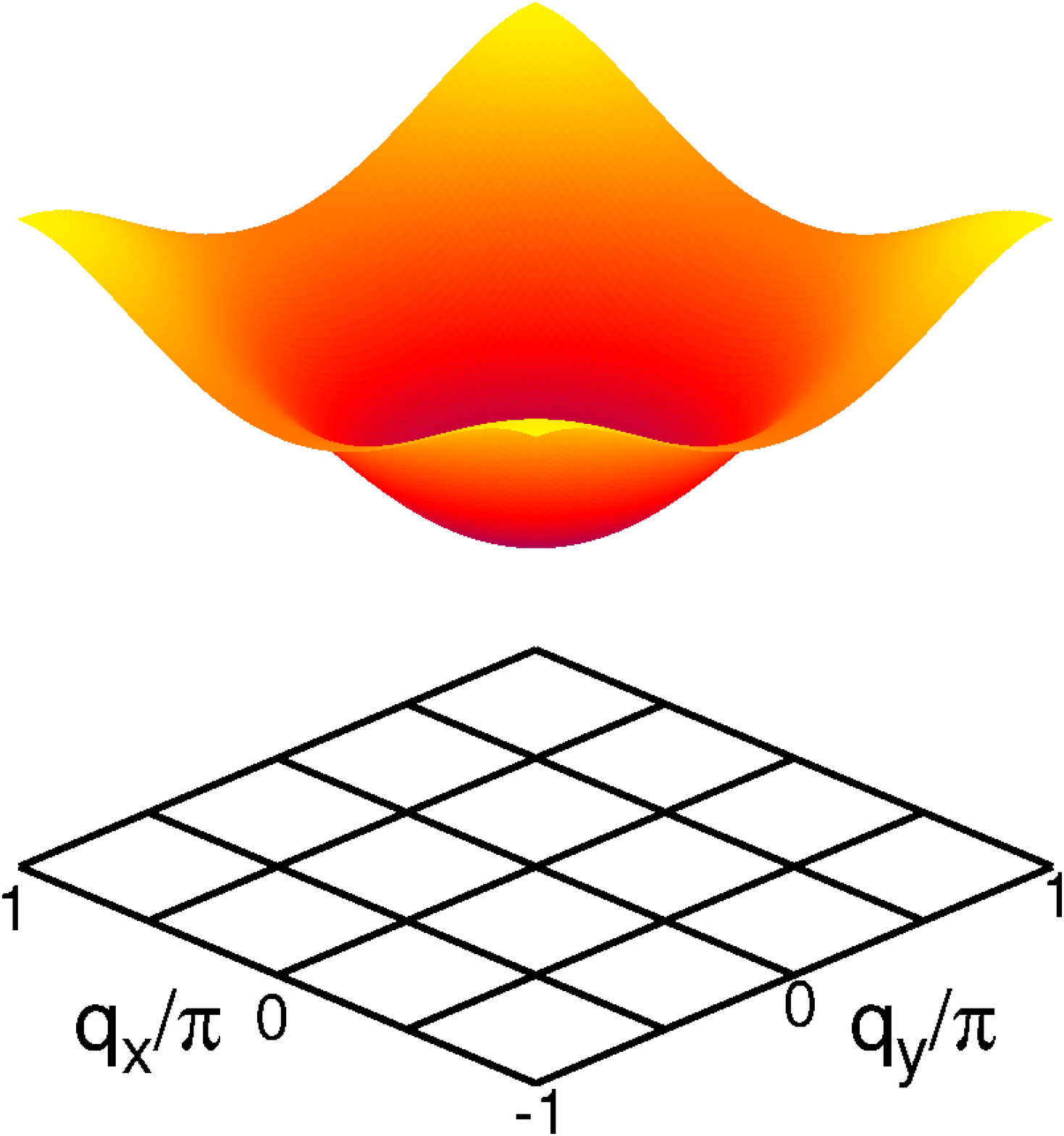}
\label{fig3f}
}
\subfigure[$U=50.0$,\newline $\widetilde{T}=1.108$]{
\includegraphics[height=0.25\linewidth]{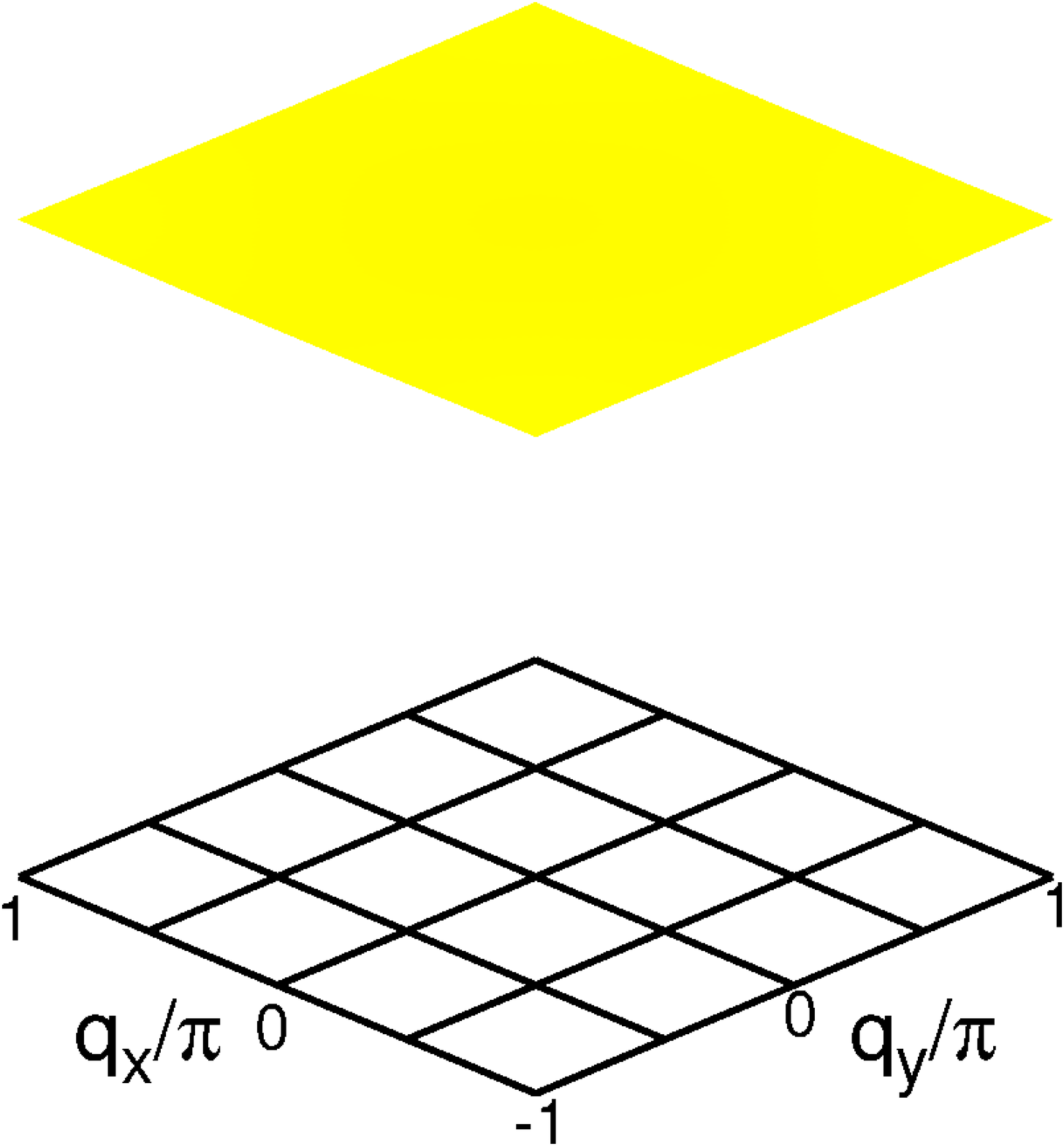}
\label{fig3g}
}
\subfigure{
\includegraphics[height=0.25\linewidth]{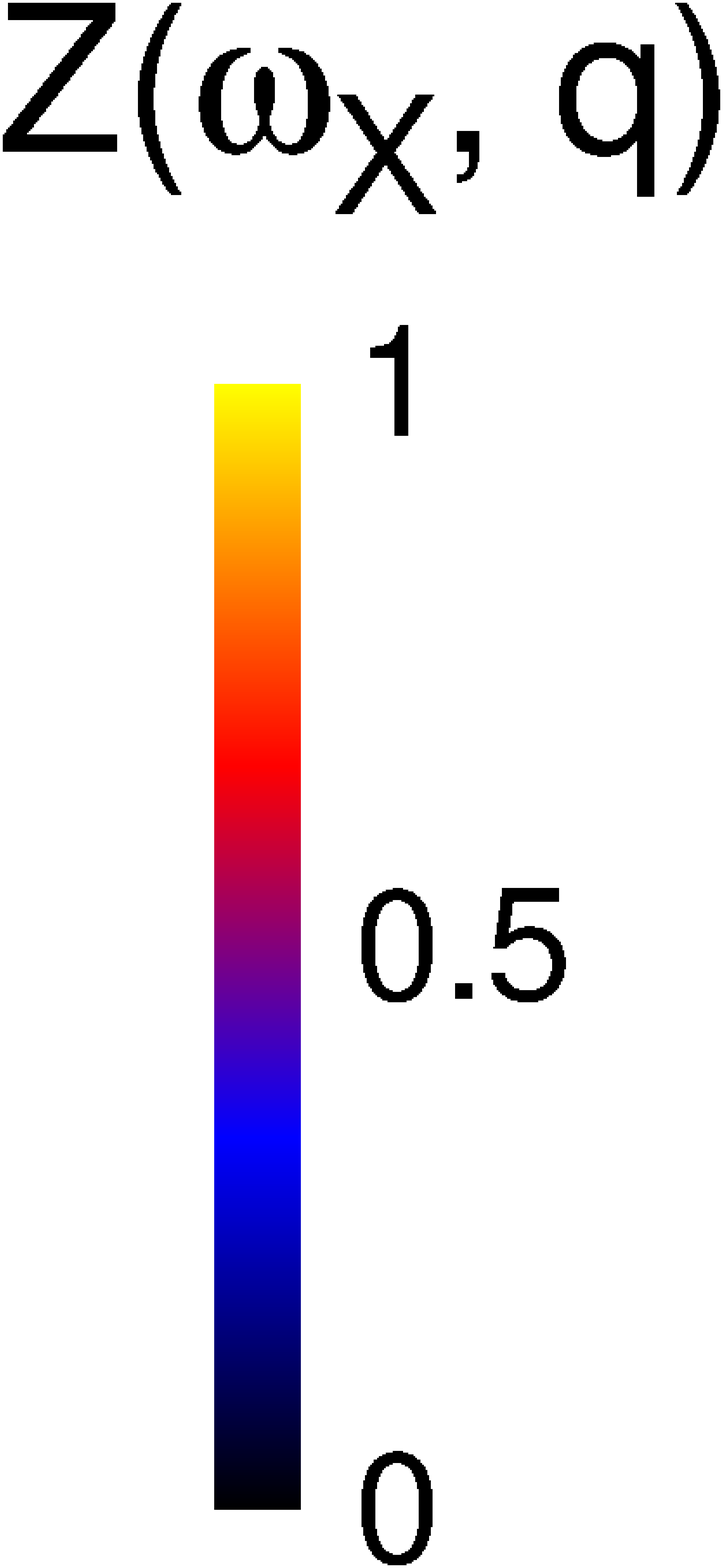}
\label{fig3h}
}
\caption{(Color online) Exciton quasiparticle 
weight $Z(\omega_X,\q)$.}
\label{fig3}
\end{figure}

Now, let us take a closer look as to how the excitonic instability develops
in the SM and SC regimes (cf. Figs.~\ref{fig1},~\ref{fig2}  and~\ref{fig3}). 
We start with the SM (point {\large \textcircled{\normalsize 1}} in 
Fig.~\ref{fig1}). Here the valence and conduction bands slightly overlap; 
as a result a distinct Fermi surface exists (see Fig.~\ref{fig2}, blue frame). 
Approaching $T_{\rm EI}$, the electron-hole pair fluctuations contained 
in $\chi_{\bf q}^{-\sigma,\sigma}$ become critical and will drive a phase 
transition, which is accompanied by a spontaneous hybridization of 
the $\uparrow$- and $\downarrow$-bands~\cite{MSGMDCMBTBA10}. 
The resulting energy spectrum exhibits a gap at the Fermi level, 
where the density-of-states is largely enhanced at the top (bottom)
of the lower (upper) quasiparticle band~\cite{IPBBF08,ZIBF10}. 
The pivotal question is whether excitons are involved in 
this BCS-like transition. While excitonic bound states might 
exist above $T_{\rm EI}$ (in the region given by the red line), we 
definitely have no excitons with ${\bf q}=0$. In either case, 
$Z(\omega_X,\q)$ is zero except near the corners of the 
Brillouin zone [see Fig.~\ref{fig3}(b)], and the number of 
these excitons, having a large center-of-mass momentum,  
is very small (see $N_X(\q)$ in Fig.~\ref{fig2}). Hence, the formation 
of the EI state in the SM region is barely influenced by excitons.

A larger Coulomb interaction $U$ will affect the system in 
two ways: It (i) increases the bare band splitting and (ii) amplifies 
the attraction between electrons and holes. At the SM-SC transition 
(point {\large \textcircled{\normalsize 2}} in Fig.~\ref{fig1}, 
purple frame in Fig.~\ref{fig2}), the $\uparrow$- and $\downarrow$ bands
only touch each other (at ${\bf k}=0$). Accordingly the Fermi surface shrinks
in size to a point. In this case a larger number of excitons form
(also with small momenta), but again the zero-momentum excitons 
play no significant role because of the cone-like structure 
of $Z(\omega_X,\q)$; see Fig.~\ref{fig3}(c).

In the SC region (point {\large \textcircled{\normalsize 3}} in 
Fig.~\ref{fig1}, green frame in Fig.~\ref{fig2}) the (Hartree) 
band structure exhibits a direct gap, within which the 
chemical potential $\mu$ and the exciton level $\omega_X$ are located. Now zero-momentum 
excitons may occur. Although having the lowest binding energy,  
they represent the largest contribution to the total number of 
excitons.  
Here $Z(\omega_X,\q)$ is finite for all momenta.
Actually the system now realizes a three-component plasma consisting
of electrons, holes, and excitons. 
Lowering the temperature, the excitonic level moves toward the 
valence band and finally touches its top at ${\bf k}=0$ 
(point {\large \textcircled{\normalsize 4}} in Fig.~\ref{fig1}, 
red frame in Fig.~\ref{fig2}). Thereby the excitonic instability appears,  
and the SC-EI transition takes place. Most notably, we 
observe a divergence of $N_X(0)$, i.e., the zero-momentum excitonic 
state is macroscopically occupied. This demonstrates the BEC of 
preformed excitons, contrary to the BCS-like transition on the SM side. 

The spectral weight $Z(\omega_X,\q)$ 
apparently accounts for the character 
and composite nature of the electron-hole bound states. This becomes 
especially evident in the weak and strong interaction limits. 
For very small $U$, 
the Coulomb attraction 
between electrons and holes can neither form excitonic bound states 
nor establish the $c$-$f$ electron coherence 
in the EI state. Here $Z(\omega_X,\q)=0$, independent of 
${\bf q}$ [see Fig.~\ref{fig3}(a)]. By contrast, as 
$U\rightarrow\infty$,  $Z(\omega_X,\q)=1$ $\forall\,\q$ 
[Fig.~\ref{fig3}(g)]. Hence, in this limit, excitons behave as ideal 
bosons; cf. Eq.~\eqref{Xnumbr}. For $U\to \infty$,  
$\omega_X(\q)$ scales as $\ln U$ while the continuum level grows 
$\propto U$ [recall that $\omega_C(0)=-E_\uparrow
-4(1+|t_\uparrow|)+U(n_\uparrow-n_\downarrow)$]; 
$E_B^X$ becomes infinite. Despite this, an EI phase cannot be 
established, this time because the large band splitting prevents
$c$-$f$ electron coherence. This explains why the EI phase arises  
below an upper critical coupling $U_c$ only. 

\begin{figure}[t]
\centering
\includegraphics[width=\linewidth]{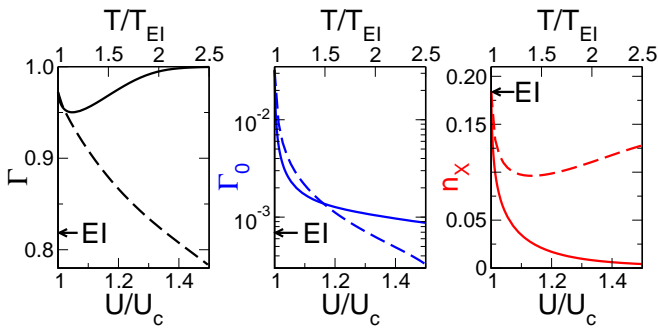}
\caption{(Color online) Bound-state fractions $\Gamma$ 
and $\Gamma_0$ as functions of Coulomb attraction $U$  at 
fixed $\widetilde{T}=0.679$ (solid lines, lower scales), 
and as  functions of temperature at fixed $U=5.5$ 
[dashed lines, upper scales; here $T$ is given in units of
$T_{\rm EI}(U=5.5)$]. The right panel shows the 
corresponding exciton densities $n_X$. 
}
\label{fig4}
\end{figure}

Having identified the nature and the condensation mechanism of 
electron-hole pairs we now discuss how they might influence the normal state 
properties of the EFKM. In particular, we examine the so-called halo phase 
around the EI, where excitons and excitonic resonances 
dominate the electron-hole excitation spectrum~\cite{BF06,PFB11}. 
Figure~\ref{fig4} gives results for the SC region with   
$U>U_c(T)$ and $T>T_{\rm EI}(U)$. Already for $U/U_c(T) \simeq 1.5$ 
almost all electron-hole pair excitations constitute excitons, 
i.e., $\Gamma\to 1$.  
The small number of nearly-free $\downarrow$-electrons can be 
attributed to the relatively large band gap. Remarkably, the portion 
of excitons with ${\bf q}=0$ is less than one per thousend.
Approaching $U_c$ from above, by reducing $U$ at fixed temperature, 
the fraction of the zero-momentum excitons increases by two orders 
of magnitude, thereby overcompensating the initial depletion of $\Gamma$ 
caused by the reduction of $U$. Thus, on the SC side
the formation of the EI is driven by the condensation of 
zero-momentum excitons. Keeping $U$ constant and coming up to the
EI by lowering the temperature, we observe an uninflected increase
of both $\Gamma$ and $\Gamma_0$ which again is triggered by the
occupation of excitonic bound-states with ${\bf q}=0$. Here the initial decrease
of $n_X$ results from the narrowing of the Bose distribution.
For temperatures of about $T/T_{\rm EI}(U) \simeq 2.5$, we 
find a significant number of unbound $\downarrow$-electrons. Their 
contribution increases, if $T$ is further 
raised, because excitons dissociate.   

\begin{figure}[t]
\centering
\includegraphics[width=\linewidth]{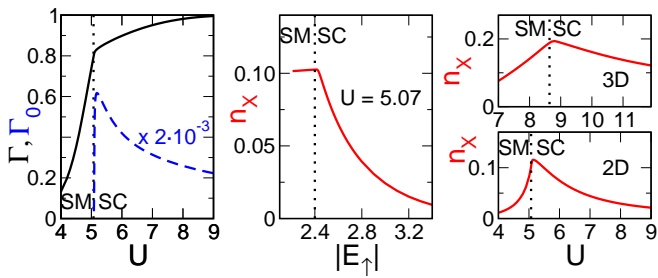}
\caption{(Color online) Bound-state fractions $\Gamma$ (solid line) and
$\Gamma_0$ (dashed line) as functions of the Coulomb attraction $U$ 
(left panel), and exciton density $n_X$ versus
bare band splitting $|E_{\uparrow}|$ (middle panel), both 
at $\widetilde{T}=1.108$.  
The right-hand panels compare the $U$-dependence of $n_X$  
for the 2D and 3D cases (at the same reduced temperature $\widetilde{T}$).
}
\label{fig5}
\end{figure}

Figure~\ref{fig5} illustrates what happens if we cross the border to the
SM phase at fixed $\tilde{T}>1$ by reducing the bare band splitting 
$|E_\uparrow|$ (middle panel) or downsizing  $U$  (right panels). 
On the SC side, $n_X$ increases because the band gap decreases and 
concomitantly the exciton level deepen.   
In the SM, zero-momentum excitons cannot exist, and $\Gamma_0$ drops to zero.
Although small, $\Gamma$ is finite nevertheless, because 
electron-hole bound states carrying a finite momentum remain.
These excitonic resonances will affect the transport properties
on the SM side as well. Thus, basically the whole EI phase is surrounded 
by an exciton-rich region (halo); there the number 
of charge carriers is substantially reduced, and  
excitons provide abundant scattering centers for the
residual electrons and holes. We expect that an inclusion 
of the continuum electron-hole scattering states will round off 
the sharp kinks appearing in Fig.~\ref{fig5} at the 
SC-SM transition~\cite{Ro94}. 

Now we relate our results to experiments on the SC-SM transition in 
$\rm TmSe_{0.45} Te_{0.55}$~\cite{NW90,BSW91,Wa01,WBM04}.
The anomaly in the lattice expansion as a function of temperature at
high constant pressure occurring near 250~K was ascribed to a SM-EI
transition, and the ratio of the exciton density $n_{\rm ex}$ and the atomic
density $n_{\rm Tm}$ was estimated as $n_{\rm ex}/n_{\rm Tm} = 0.22$
in Ref.~\onlinecite{Wa01}. Assuming
Mott-Wannier-type excitons, they are suggested to overlap due to their
large concentration. However, the binding energy was found to be too
large~\cite{NW90}, which questions the Mott-Wannier-type model.  In our EFKM
model, the coherence length $r_{\rm coh}$ of the excitons  
at $T = 0$ and for $U\sim 3.6$ equals the lattice 
constant (see Fig.~\ref{fig1} and Ref.~\onlinecite{SEO11}). 
At the SC-EI transition at $T_{\rm EI}(U = 5.5)$ we obtain $n_X \sim
0.18$ (see Fig.~\ref{fig4}, right panel), and a BEC of
non-overlapping Frenkel-type excitons with a high density takes place.
The numerical value of $n_X$ at the SC side of the phase diagram 
approximately agrees with the experimental value [note that 
the agreement improves for the (real) 3D situation, 
see Fig~\ref{fig5}, right upper panel]. 
Taking the $f$ bandwidth $W_{\uparrow} = 8 |t_{\uparrow}| \simeq 30$~meV
(Ref.~\onlinecite{Wa01}) and our parameter choice, we get 
$T_{\rm EI}^{\rm max}  \simeq 0.3 t_{\downarrow} \simeq 20$~K. That means, the
experimental phase boundary at the SC side between 20~K and 250~K
obtained by electrical resisitivity data~\cite{NW90,BSW91,Wa01,WBM04}
describes the appearance of an exciton-rich halo phase above the SC-EI
transition, as was also concluded in Ref.~\onlinecite{BF06}. On the other hand,
the observed linear increase of the heat conductivity and thermal
diffusivity with decreasing temperature below 20~K (Ref.~\onlinecite{WBM04}) 
may be ascribed to the EI phase. As revealed by measurements of the Hall
constant at 4.2~K as function of pressure~\cite{BSW91}, the position of the
maximum in the resistivity coincides with that of the minimum in the
current-carrier density. We suggest that this close relation,
indicating the formation of excitons from free current carriers, also
holds in the halo phase. Then the maximum in $n_X$ at the SC-SM
transition (see Fig.~\ref{fig5}) should correspond to a minimum in the
current-carrier density, so that the resistivity maximum at the
pressure-induced SC-SM transition in $\rm Tm Se_{0.45} Te_{0.55}$ may be
qualitatively understood within our halo-phase concept.

In summary, we have analyzed the formation of the EI  
state at the SM-SC transition in  the 
2D EFKM and provided strong evidence for a BCS-BEC crossover scenario. 
While  Cooper-type pairing fluctuations 
become critical on the SM side, Bose condensation of 
preformed zero-momentum excitons takes place on the SC side.
Accordingly, the surroundings of the EI are dominated by 
electron-hole fluctuations or excitonic bound-states
with strong impact on the transport and optical
properties. 

We thank K. W. Becker and V.-N. Phan for valuable discussions. 
This work supported by DFG, SFB 652.

\begin{thebibliography}{22}
\expandafter\ifx\csname natexlab\endcsname\relax\def\natexlab#1{#1}\fi
\expandafter\ifx\csname bibnamefont\endcsname\relax
  \def\bibnamefont#1{#1}\fi
\expandafter\ifx\csname bibfnamefont\endcsname\relax
  \def\bibfnamefont#1{#1}\fi
\expandafter\ifx\csname citenamefont\endcsname\relax
  \def\citenamefont#1{#1}\fi
\expandafter\ifx\csname url\endcsname\relax
  \def\url#1{\texttt{#1}}\fi
\expandafter\ifx\csname urlprefix\endcsname\relax\def\urlprefix{URL }\fi
\providecommand{\bibinfo}[2]{#2}
\providecommand{\eprint}[2][]{\url{#2}}




\bibitem[{\citenamefont{Halperin and Rice}(1968)}]{HR68b}
\bibinfo{author}{\bibfnamefont{B.~I.} \bibnamefont{Halperin}} \bibnamefont{and}
  \bibinfo{author}{\bibfnamefont{T.~M.} \bibnamefont{Rice}},
  \bibinfo{journal}{Rev. Mod. Phys.} \textbf{\bibinfo{volume}{40}},
  \bibinfo{pages}{755} (\bibinfo{year}{1968}).

\bibitem[{\citenamefont{Bronold and Fehske}(2006)}]{BF06}
\bibinfo{author}{\bibfnamefont{F.~X.} \bibnamefont{Bronold}} \bibnamefont{and}
  \bibinfo{author}{\bibfnamefont{H.}~\bibnamefont{Fehske}},
  \bibinfo{journal}{Phys. Rev. B} \textbf{\bibinfo{volume}{74}},
  \bibinfo{pages}{165107} (\bibinfo{year}{2006}).

\bibitem[{\citenamefont{Ihle et~al.}(2008)\citenamefont{Ihle, Pfafferott,
  Burovski, Bronold, and Fehske}}]{IPBBF08}
\bibinfo{author}{\bibfnamefont{D.}~\bibnamefont{Ihle}},
  \bibinfo{author}{\bibfnamefont{M.}~\bibnamefont{Pfafferott}},
  \bibinfo{author}{\bibfnamefont{E.}~\bibnamefont{Burovski}},
 \bibinfo{author}{\bibfnamefont{F.~X.} \bibnamefont{Bronold}},
  \bibnamefont{and} \bibinfo{author}{\bibfnamefont{H.}~\bibnamefont{Fehske}},
  \bibinfo{journal}{Phys. Rev. B} \textbf{\bibinfo{volume}{78}},
  \bibinfo{pages}{193103} (\bibinfo{year}{2008}).

\bibitem[{\citenamefont{Seki et~al.}(2011)\citenamefont{Seki, Eder, and
  Ohta}}]{SEO11}
\bibinfo{author}{\bibfnamefont{K.}~\bibnamefont{Seki}},
  \bibinfo{author}{\bibfnamefont{R.}~\bibnamefont{Eder}}, \bibnamefont{and}
  \bibinfo{author}{\bibfnamefont{Y.}~\bibnamefont{Ohta}}, 
  \bibinfo{journal}{Phys. Rev. B} \textbf{\bibinfo{volume}{84}},
  \bibinfo{pages}{245106} (\bibinfo{year}{2011}).

\bibitem[{\citenamefont{Bu}(1990)}]{BSW91}
B. Bucher, P. Steiner, and P. Wachter, Phys. Rev. Lett. {\bf 67}, 2717 (1991).

\bibitem[{\citenamefont{Neuenschwander and Wachter}(1990)}]{NW90}
\bibinfo{author}{\bibfnamefont{J.}~\bibnamefont{Neuenschwander}}
  \bibnamefont{and} \bibinfo{author}{\bibfnamefont{P.}~\bibnamefont{Wachter}},
  \bibinfo{journal}{Phys. Rev. B} \textbf{\bibinfo{volume}{41}},
  \bibinfo{pages}{12693} (\bibinfo{year}{1990}).

\bibitem[{\citenamefont{Wachter}(2001)}]{Wa01}
P. Wachter, Solid State Commun. {\bf 118}, 645 (2001).

\bibitem[{\citenamefont{Wachter et~al.}(2004)\citenamefont{Wachter, Bucher, and
  Malar}}]{WBM04}
\bibinfo{author}{\bibfnamefont{P.}~\bibnamefont{Wachter}},
  \bibinfo{author}{\bibfnamefont{B.}~\bibnamefont{Bucher}}, \bibnamefont{and}
  \bibinfo{author}{\bibfnamefont{J.}~\bibnamefont{Malar}},
  \bibinfo{journal}{Phys. Rev. B} \textbf{\bibinfo{volume}{69}},
  \bibinfo{pages}{094502} (\bibinfo{year}{2004}).



\bibitem{Wa09} Y. Wakisaka, T. Sudayama, K. Takubo, T. Mizokawa, M. Arita, H. Namatame, M. Taniguchi, N. Katayama, M. Nohara, and H. Takagi, 
Phys. Rev. Lett {\bf 103}, 026402 (2009);
T. Kanaeko, T. Toriyama, Y. Ohta, and T. Konishi, (2011), preprint.

\bibitem[{\citenamefont{Monney et~al.}(2010)\citenamefont{Monney, Schwier,
  Garnier, Mariotti, Didiot, Cercellier, Marcus, Berger, Titov, Beck
  et~al.}}]{MSGMDCMBTBA10}
\bibinfo{author}{\bibfnamefont{C.}~\bibnamefont{Monney}}
  \bibinfo{author}{\bibfnamefont{E.~F.} \bibnamefont{Schwier}},
  \bibinfo{author}{\bibfnamefont{M.~G.} \bibnamefont{Garnier}},
  \bibinfo{author}{\bibfnamefont{N.}~\bibnamefont{Mariotti}},
  \bibinfo{author}{\bibfnamefont{C.}~\bibnamefont{Didiot}},
  \bibinfo{author}{\bibfnamefont{H.}~\bibnamefont{Cercellier}},
  \bibinfo{author}{\bibfnamefont{J.}~\bibnamefont{Marcus}},
  \bibinfo{author}{\bibfnamefont{H.}~\bibnamefont{Berger}},
  \bibinfo{author}{\bibfnamefont{A.~N.} \bibnamefont{Titov}},
  \bibinfo{author}{\bibfnamefont{H.}~\bibnamefont{Beck}}, \bibnamefont{and}
 \bibinfo{author}{\bibfnamefont{P.}~\bibnamefont{Aebi}}
  \bibinfo{journal}{New J. Phys.} \textbf{\bibinfo{volume}{12}},
  \bibinfo{pages}{125019} (\bibinfo{year}{2010});
\bibinfo{author}{\bibfnamefont{C.}~\bibnamefont{Monney}},
  \bibinfo{author}{\bibfnamefont{C.}~\bibnamefont{Battaglia}},
  \bibinfo{author}{\bibfnamefont{H.}~\bibnamefont{Cercellier}},
  \bibinfo{author}{\bibfnamefont{P.}~\bibnamefont{Aebi}}, \bibnamefont{and}
 \bibinfo{author}{\bibfnamefont{H.}~\bibnamefont{Beck}},
  \bibinfo{journal}{Phys. Rev. Lett.} \textbf{\bibinfo{volume}{106}},
  \bibinfo{pages}{106404} (\bibinfo{year}{2011}).




\bibitem[{\citenamefont{Lozovik and Sokolik}(2008)}]{LS08b}
\bibinfo{author}{\bibfnamefont{Y.~E.} \bibnamefont{Lozovik}} \bibnamefont{and}
  \bibinfo{author}{\bibfnamefont{A.~A.} \bibnamefont{Sokolik}},
  \bibinfo{journal}{JETP Letters} \textbf{\bibinfo{volume}{87}},
  \bibinfo{pages}{55} (\bibinfo{year}{2008});
H. Min, R. Bistritzer, J.-J. Su, and A. H. MacDonald,
Phys. Rev. B {\bf 78}, 121401 (2008); 
V.-N. Phan and H. Fehske, arXiv:1202.0900.

\bibitem[{\citenamefont{Leggett}(1980)}]{Le80}
\bibinfo{author}{\bibfnamefont{A.~J.} \bibnamefont{Leggett}}, in
  \emph{\bibinfo{booktitle}{Modern Trends in the Theory of Condensed Matter}},
  edited by \bibinfo{editor}{\bibfnamefont{A.}~\bibnamefont{Pekalski}}
  \bibnamefont{and} \bibinfo{editor}{\bibfnamefont{R.}~\bibnamefont{Przystawa}}
  (\bibinfo{publisher}{Springer-Verlag}, \bibinfo{address}{Berlin},
  \bibinfo{year}{1980});
\bibinfo{author}{\bibfnamefont{C.}~\bibnamefont{Comte}} \bibnamefont{and}
  \bibinfo{author}{\bibfnamefont{P.}~\bibnamefont{Nozi\`eres}},
  \bibinfo{journal}{J. Phys. (France)} \textbf{\bibinfo{volume}{43}},
  \bibinfo{pages}{1069} (\bibinfo{year}{1982});
\bibinfo{author}{\bibfnamefont{P.}~\bibnamefont{Nozi\`eres}} \bibnamefont{and}
 \bibinfo{author}{\bibfnamefont{S.}~\bibnamefont{Schmitt-Rink}},
 \bibinfo{journal}{J. Low Temp. Phys.} \textbf{\bibinfo{volume}{59}},
 \bibinfo{pages}{195} (\bibinfo{year}{1985}).

\bibitem[{\citenamefont{Batista}(2002)}]{Ba02b}
\bibinfo{author}{\bibfnamefont{C.~D.} \bibnamefont{Batista}},
  \bibinfo{journal}{Phys. Rev. Lett.} \textbf{\bibinfo{volume}{89}},
  \bibinfo{pages}{166403} (\bibinfo{year}{2002});
\bibinfo{author}{\bibfnamefont{C.~D.} \bibnamefont{Batista}},
  \bibinfo{author}{\bibfnamefont{J.~E.} \bibnamefont{Gubernatis}},
  \bibinfo{author}{\bibfnamefont{J.}~\bibnamefont{Bon\v{c}a}},
  \bibnamefont{and} \bibinfo{author}{\bibfnamefont{H.~Q.} \bibnamefont{Lin}},
  \bibinfo{journal}{Phys. Rev. Lett.} \textbf{\bibinfo{volume}{92}},
  \bibinfo{pages}{187601} (\bibinfo{year}{2004}).

\bibitem[{\citenamefont{Farka\v{s}ovsk\'{y}}(2008)}]{Fa08}
\bibinfo{author}{\bibfnamefont{P.}~\bibnamefont{Farka\v{s}ovsk\'{y}}},
  \bibinfo{journal}{Phys. Rev. B} \textbf{\bibinfo{volume}{77}},
  \bibinfo{pages}{155130} (\bibinfo{year}{2008});
\bibinfo{author}{\bibfnamefont{C.}~\bibnamefont{Schneider}} \bibnamefont{and}
  \bibinfo{author}{\bibfnamefont{G.}~\bibnamefont{Czycholl}},
  \bibinfo{journal}{Eur. Phys. J. B} \textbf{\bibinfo{volume}{64}},
  \bibinfo{pages}{43} (\bibinfo{year}{2008}).

\bibitem[{\citenamefont{Zenker et~al.}(2010)\citenamefont{Zenker, Ihle,
  Bronold, and Fehske}}]{ZIBF10}
\bibinfo{author}{\bibfnamefont{B.}~\bibnamefont{Zenker}},
  \bibinfo{author}{\bibfnamefont{D.}~\bibnamefont{Ihle}},
  \bibinfo{author}{\bibfnamefont{F.~X.} \bibnamefont{Bronold}},
  \bibnamefont{and} \bibinfo{author}{\bibfnamefont{H.}~\bibnamefont{Fehske}},
  \bibinfo{journal}{Phys. Rev. B} \textbf{\bibinfo{volume}{81}},
  \bibinfo{pages}{115122} (\bibinfo{year}{2010});
%
{\it ibid.} \textbf{\bibinfo{volume}{83}},
  \bibinfo{pages}{235123} (\bibinfo{year}{2011}).

\bibitem[{\citenamefont{Plakida}(2011)}]{Pl11}
\bibinfo{author}{\bibfnamefont{N.~M.} \bibnamefont{Plakida}},
  \bibinfo{journal}{Springer Series in Solid-State Sciences}
  \textbf{\bibinfo{volume}{171}}, \bibinfo{pages}{173} (\bibinfo{year}{2011}).

\bibitem[{\citenamefont{Phan et~al.}(2011)\citenamefont{Phan, Becker, and
  Fehske}}]{PFB11}
\bibinfo{author}{\bibfnamefont{V.-N.} \bibnamefont{Phan}},
  \bibinfo{author}{\bibfnamefont{K.~W.} \bibnamefont{Becker}},
  \bibnamefont{and} \bibinfo{author}{\bibfnamefont{H.}~\bibnamefont{Fehske}},
  \bibinfo{journal}{Europhys. Lett.} \textbf{\bibinfo{volume}{95}},
  \bibinfo{pages}{17006} (\bibinfo{year}{2011}).

\bibitem{Ro94}
G.~R\"opke, Ann. Physik {\bf 3}, 145 (1994).

\end{thebibliography}

\end{document}